\documentclass [a4paper, 11pt]{article}

\usepackage{amsmath,amssymb}
\usepackage[english]{babel}
\usepackage{color}
\usepackage{float}
\usepackage[margin=33mm]{geometry}
\usepackage{subfig}
\usepackage{graphicx}
\usepackage{xspace}
\usepackage{url}
\usepackage[misc]{ifsym}

\newcommand{\hypre}{hypre\xspace}
\newcommand{\juqueen}{Juqueen\xspace}
\newcommand{\parflow}{ParFlow\xspace}
\newcommand{\pforest}{\texttt{p4est}\xspace}

\newcommand{\citep}[1]{\cite{#1}}

\title{Enhancing speed and scalability of the \parflow simulation code}

\author{Carsten Burstedde \and Jose A.\ Fonseca \and Stefan Kollet}

\author{%
Carsten Burstedde\thanks{Institut f\"ur Numerische Simulation (INS)
and Hausdorff Center for Mathematics (HCM), Rheini\-sche Friedrich-Wilhelms-Universit\"at Bonn, Germany.}, %
Jose A.\ Fonseca\footnotemark[1] 
\footnote{Corresponding author: \texttt{fonseca@ins.uni-bonn.de}}, %
Stefan Kollet\thanks{Agrosphere (IGB-3), Forschungszentrum J\"ulich GmbH, and
Centre for High-Performance Scientific Computing in Terrestrial Systems, Geoverbund ABC/J, J\"ulich, Germany.}
}
\date{}

\begin{document}

\maketitle

\begin{abstract}
Regional hydrology studies are often supported by high resolution
simulations of subsurface flow that require expensive and extensive
computations.
Efficient usage of the latest high performance parallel computing systems
becomes a necessity.
The simulation software \parflow has been demonstrated to meet this requirement
and shown to have excellent solver scalability for up to 16,384 processes.

In the present work we show that the code requires further enhancements in
order to ful\-ly take advantage of current petascale machines.
We identify \parflow's way of parallelization of the computational mesh as a
central bottleneck.
We propose to reorganize this subsystem using fast mesh partition algorithms
provided by the parallel adaptive mesh refinement library \pforest.
We realize this in a minimally invasive manner by modifying selected parts of
the code to reinterpret the existing mesh data structures.
%
We evaluate the scaling performance of the modified version of \parflow,
demonstrating good weak and strong scaling up to 458k cores of the Ju\-queen
supercomputer, and test an example application at large scale.



\end{abstract}


\section{Introduction}

The accurate simulation of variably saturated flow in porous media is a
valuable component to understanding physical processes occurring in many water
resource problems.
Examples range from coupled hydrologic-at\-mospheric models to irrigation
problems; see for example \cite{YamamotoZhangKarasakiEtAl09,CamporesePaniconiPuttiEtAl10,KuznetsovYakirevichPachepskyEtal12}.
Computer simulations of subsurface flow generally proceed by computing a
numerical solution of the three dimensional Richards' equation
\cite{Richards31}.
Assuming a two-phase water-gas system, Richards' equation can be derived from
the generalized Darcy laws \cite{Muskat81} under the assumption that the
pressure gradient in the gas phase is small.
One of the variants of Richards' equation reads
\begin{subequations}
  \label{richards}
\begin{gather}
  \dfrac{\partial (\phi s(p))}{\partial t} + \nabla\cdot \vec u = f, \\
  \vec u = -K \nabla (p - z),
\end{gather}
\end{subequations}
where $p$ denotes the pressure head, $s(p)$ is the pressure-dependent saturation,
$\phi$ the porosity of the medium, $\vec u$ the flux,
$z$ is the depth below the surface, $K = K(x; p)$ the symmetric conductivity
tensor and $f$ the source term.

The numerical solution of \eqref{richards} is challenging because of two main
reasons.
The first is the nonlinearity and large variation in the equation's
coefficients \cite{JonesWoodward01}, essentially introduced by the pressure
dependent conductivity tensor.
The second is the requirement of discretizing very large temporal and spatial
domains with a resolution sufficient for detailed physics based
hydrological models \cite{KolletMaxwellWoodwardEtAl10}.
As a consequence, demands for computational time and memory resources for
computer simulations of subsurface flow are enormous, and considerations of
efficiency become prominent.
As pointed out by \cite{MillerClintDawsonEtAl13},
a clear computational trend is that the computational power is provided by
parallel computing.
Hence, effective employment of high performance (parallel) computing strategies
(HPC) plays a key role in solving water resource problems.

There has been a significant focus across the hydrology modeling community
to incorporate modern HPC paradigms into their simulators, see for example
\cite{HardelaufJavauxHerbstEtAl07,ZhangwuPruess08,HammondLichtnerMills14,HwangParkSudickyEtAl14,OrgogozoRenonSoulaineEtAl14,Liu13}.
We provide just a short and necessarily incomplete summary.
\begin{enumerate}
  \item
        PARSWMS \cite{HardelaufJavauxHerbstEtAl07} is an MPI-parallelized code
        written in C++.
        The finite element (FE) discretization relies on an unstructured mesh
        managed by the ParMETIS library \cite{KarypisKumar98}.
        The solvers are provided by the PETSc library
        \cite{BalayBrownBuschelmanEtAl12}.
        Strong scaling studies have been performed for a problem with 492k
        degrees of freedom (dofs) and process counts between 1 and 256.
  \item
        TOUGH2-MP \cite{ZhangwuPruess08} is a Fortran/MPI code.
        It uses an integral finite difference (FD) discretization on an
        unstructured mesh, METIS partitioning \cite{KarypisKumar95} and the
        Aztec linear solver \cite{TuminaroHerouxHutchinsonEtAl99}.
        Strong scaling studies have been published for 1--256 cores.
  \item
        PFLOTRAN \cite{HammondLichtnerMills14} is an MPI code written in free
        format Fortran 2003.
        The parallel solvers and the interface to METIS are provided by PETSc.
        Weak and strong scaling studies are available up to 16,384 processes.
        The biggest problem treated has $10^{7}$ dofs.
        Scalability is good when the number of dofs per core is at least 10k.
  \item
        Hydrogeosphere HGS \cite{HwangParkSudickyEtAl14} is an OpenMP code.
        An unstructured mesh is the basis for a fully implicit discretization
        via the control volume finite element method, a combination of centered
        FD and FE.
        For parallel computation, the domain is partitioned into subdomains,
        and a multiblock node reordering is executed.
        Strong scaling studies over up to 16 threads for a problem with $10^{7}$
        unknowns have been reported.
  \item
        RichardsFOAM \cite{OrgogozoRenonSoulaineEtAl14} and suGWFOAM
        \cite{Liu13} are co\-des based on OpenFOAM \cite{OpenCFD07}.
        They inherit its MPI parallelism and use a finite volume (FV)
        discretization on an unstructured mesh.
        The mesh is partitioned with METIS.
        Scaling studies refer to 1--1024 processes for problems on the order
        between $2\times10^6$ and $133\times10^6$ cells.
  \item
        \parflow
        \cite{AshbyFalgout96,JonesWoodward01,KolletMaxwell06,KolletMaxwellWoodwardEtAl10} is a
	MPI-parallelized code mainly written in C. It uses a FD discretization
	on a structured Cartesian mesh. Nonlinear solvers and preconditioners
	are provided by the KINSOL \cite{HindmarshBrownGrantEtAl05} and
	\hypre \cite{HypreTeam12} packages, respectively.
	Weak scaling studies are available up to 16,384 processes. The biggest
	problem reported has $8\times10^9$ dofs.
\end{enumerate}


In line with the current state of the practice summarized above,
we henceforth consider a parallel computer that implements the MPI standard.
It consists of multiple physical compute nodes connected by a network.
Each node has access to the memory physically located in that node,
thus we speak of distributed memory and distributed parallelization.
A node has multiple central processing units (CPUs), consisting of one or
more CPU cores each, with each core running one or more processes or threads.
For the purpose of this discussion, we will use the terms process, CPU core,
and CPU interchangeably, really referring to one MPI process as the atomic unit
of parallelization.

The ideal hypothesis of parallel computing is that subdividing a task fairly
among several processes will result in a proportional reduction of the overall
runtime.
In order to effectively produce such behavior, parallel codes should meet two
basic requirements: maintain a balanced work-load per process and minimize
process-intercommunication, both in terms of the number of messages and the
message sizes.
The first criterion can be met fairly easily for uniform meshes (consider, for
example, a checkerboard-grid in 2D).
The number of messages to be sent and received depends on the exact
assignment of the mesh elements to processes:
Two different assignments for the same global mesh topology can lead to
significantly different communication volumes.
One guideline that helps bounding the communication is to make sure that each
process has only a constant number of other processes to communicate with,
independent of the size and shape of the mesh elements and the total number of
processes that we shall call $P$.
Hence, the primary item to look for when auditing a parallel code for
scalability is the size of loops over process indices:
If there are loops that iterate over all $P$ of them, such a construction may
slow down the program in the limit of many processes, quite possibly to the
point of uselessness.

Once the communication pattern is established, that is, it has been determined
which process sends a message to which, the impact of sending and receiving the
messages can be reduced by performing the communication in a background process
and organizing the program such that useful computations are carried
out while the messages are in transit.
The MPI standard supports this design by providing routines for non-blocking
communication, and most modern codes use them in one way or another to good
effect.

%
%

In the simulation platform \parflow, distributed parallelism is exploited
by subdividing the computational mesh into non-overlapping Cartesian blocks
called subgrids and identifying each of them with a unique process
in the parallel machine.
Hence, a subgrid constitutes the atomic unit of parallelization in \parflow.
They are logically arranged in a lexicographic ordering, which allows for
mathematically simple formulas to identify the indices of processes that any
given process communicates with.
It should be noted that communication between two processes requires symmetric
information, at least in the established version of MPI:
The sender must know the receiver's process index and the receiver must know
the sender's, and both must know the size of the message.
When \parflow precomputes such information,
it analyses the computational procedure defined by the
choices on numerical discretization and solvers.
It determines which of the (up to 26) neighbor subgrids of any subgrid
are relevant, which translates into the corresponding process indices.

The current implementation of this so-called setup phase utilizes loops that
iterate over the full size of the parallel machine and perform significant work
in each iteration.
As mentioned above, \parflow's way of subdividing the mesh enforces that the
number of subgrid used to split the mesh must match the process count of the
parallel machine.
Our hypothesis is that we can enhance the parallel scalability of \parflow by
reorganizing its mesh management in a way that drops the latter restriction
and, even more importantly, replaces the loops over $P$ with constant size
loops.
The challenge is to determine how exactly this can be achieved.

We propose to perform such reorganization using fast mesh refinement and
partition algorithms implemented in the parallel adaptive software library
\pforest \cite{BursteddeWilcoxGhattas11,IsaacBursteddeWilcoxEtAl15}.
It is known for its modularity and proven parallel performance
\cite{BursteddeGhattasGurnisEtAl10,RudiMalossiIsaacEtAl15,MullerKoperaMarrasEtAl15},
which derives from a strict minimalism when it comes to identifying processes
to communicate with.
We couple the \parflow and \pforest libraries such that \pforest becomes
\parflow's mesh manager.
Our approach is to identify each atomic mesh unit of \pforest with a subgrid,
changing \parflow's concept of parallel ownership to the one defined by
\pforest.
In essence we abandon the lexicographic ordering of subgrids in favor of using
a space filling curve, which opens up the potential to generalize from one to
several subgrids per process on the one hand, and from uniform to adaptive
refinement on the other.

In this work we describe the coupling between \parflow and \pforest in detail.
We refer to the product resulting from this coupling and further optimizations
as the modified version of \parflow.
We demonstrate its parallel performance by performing weak and strong scaling
studies on \juqueen \cite{Juqueen}, a Blue Gene/Q supercomputer
\cite{HaringOhmachtFoxEtAl12} that has over 458,000 CPU cores.
In comparison with the upstream version of \parflow, in which the runtime
of the mesh setup grows linearly with the number of processes, we reduce this
time by orders of magnitude (from between 10 and 40~minutes at 32K processes to
about three seconds).
%
%
In addition, a corresponding reduction in memory usage increases the value of
$P$ that can be used in practice to 458k.
Our modifications are released as open source and available to the public.

\section{The \parflow simulation platform}


In this section we present the upstream version of \parflow, which is in
widespread use and taken as the starting point for our modifications.
As mentioned above, \parflow is a simulator software for three-dimensional
variably saturated groundwater flow that is built
to exploit distributed parallelism and suited to solve large scale, high
resolution problems.
\parflow provides a solver for the three dimensional
Richards equation based on a cell centered finite difference
(FD) scheme.
It represents the update formula for each time step as a system of algebraic
equations that is solved by a Newton-Krylov nonlinear solver
\cite{JonesWoodward01}.
To reduce the number of iterations, \parflow
employs a multigrid preconditioned conjugate gradient solver
\cite{AshbyFalgout96}.
The code development has been ongoing for more that 15 years, during which time
additional features and capabilities have been added.
For example, the code has been coupled with the Common Land Model
\cite{DaiZengDickinsonEtAl03} to incorporate physical processes at the land
surface \cite{KolletMaxwell08}, and a terrain following mesh formulation has
been implemented \cite{Maxwell13} that allows \parflow to better handle
problems with fine space discretization near the ground surface.
The solver and preconditioner setup has been improved as well
\cite{Osei-KuffuorMaxwellWoodward14}.

Our main focus is on the mesh management and its parallel aspects, and how its
upstream implementation enables but also limits overall scalability.
Hence we begin by pointing out some key observations about this subsystem.


\subsection{Mesh management}

\parflow's computational mesh is uniform in all three dimensions.
The count and the spacing of mesh points in each dimension is determined by
the user via a \texttt{tcl} configuration script.  The script is loaded
at runtime and contains all parameters necessary to define
a simulation.

\parflow's mesh is logically partitioned into non-overlapping
Cartesian blocks called subgrids. The routine that allocates
a new grid essentially performs a loop over all processes in the
parallel machine and
creates a single subgrid per iteration. The parameters defining
a freshly allocated subgrid are determined by the following arithmetic.

\begin{figure}
	\centering
	\includegraphics[width=0.6\textwidth]{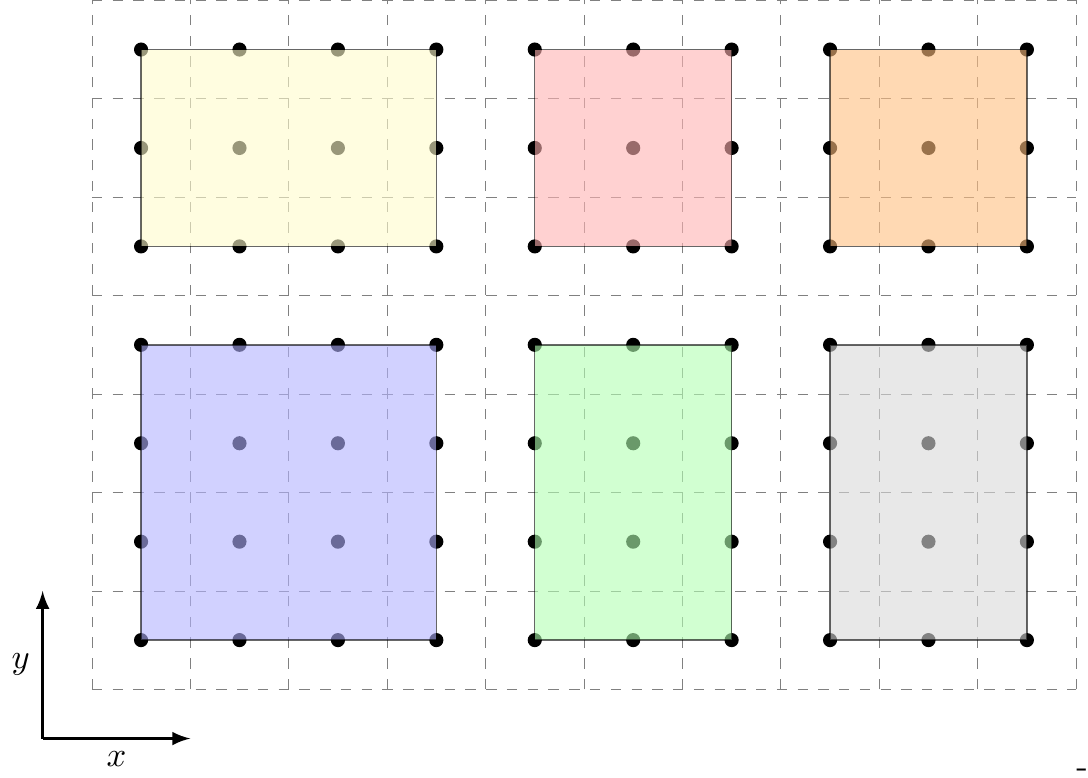}
	\caption{Example of a \parflow mesh with
		$N_x=10$ and $N_y=7$. We take $N_z=1$ to create
		a two dimensional mesh, thus the value $P_z = 1$
		is implicit. The number of processes
		is $P_x=3$, $P_y=2$, $P = 6$.
		Each shaded box is a subgrid,
		whose color symbolizes its assignment to a
		specific process.
              }
	\label{pf_subgrids}
\end{figure}

Let $P_t$ denote the number of process divisions in the $t$ coordinate
direction, for $t\in\{x,y,z\}$.
These three values are read from the script.
The total number of processes must match their product
\begin{equation}
  \label{totalprocproduct}
  P = P_x P_y P_z
  .
\end{equation}
The number of mesh points in each direction is configured in the script as
$N_{t}$ and split among $P_t$ subgrid extents as
\begin{equation}
	\label{sgdist:a}
	N_t = m_t \cdot P_t + l_t,
	\quad
	m_t\in\mathbb{N},
        \quad
        l_t\in\{0, \ldots ,P_t-1\}
	,
\end{equation}
where $m_{t}$ and $l_{t}$ are uniquely determined by
$N_{t}$ and $P_{t}$ according to
\begin{equation}
	\label{sgdist:b}
	m_{t} := N_{t} / P_{t},
        \qquad
	l_{t} := N_{t} \% P_{t}
	.
\end{equation}
Here $a/b$ denotes integer division and $a\%b$ the integer residual from
dividing $a$ by $b$.
Both $N_t$ and $P_t$ are defined by the user
in the \texttt{tcl} configuration script,
required to respect the constraint \eqref{totalprocproduct}.
Now, if $p_t$ is a process number in the range
$\{0,...,P_t-1\}$
and the triple
$p = (p_x, p_y, p_z)$
determines an index into the three dimensional process grid,
define
\begin{equation}
	\label{sgdist:corner}
	c(p_t) := p_t \cdot m_t + \min(p_t,l_t)
	,
\end{equation}
\begin{equation}
	\label{sgdist:mt}
	q(p_t) :=
	\begin{cases}
                m_t + 1 \quad &\text{if $p_t < l_t$,} \\
                m_t           &\text{otherwise.}
	\end{cases}
\end{equation}
With these definitions, the subgrid corresponding to $p$
\begin{enumerate}
\item
      has the grid point $(c(p_x),c(p_y),c(p_z))$ in its lower left
      corner,
\item has $q(p_t)$ grid points in the $t$ coordinate direction,
\item is owned by process $P_{\mathrm{own}}(p) =
        P_{\mathrm{own}}(p_x,p_y,p_z)$, where
	\begin{equation}
		\label{lexic}
		P_{\mathrm{own}}(p_x,p_y,p_z) :=
		(p_z \cdot P_y + p_y) \cdot P_x + p_x
                \in \{ 0, \ldots, P - 1 \}
		.
	\end{equation}
\end{enumerate}
An example of such a distribution of subgrids is shown
in Figure~\ref{pf_subgrids}.
It becomes clear from this logic that \parflow's computational mesh is
distributed in parallel by assigning each of
its subgrids to exactly one process via the rule \eqref{lexic}.
This order of subgrids and processes is called lexicographic.

Following the parallel distribution of the mesh, vectors and matrices in
\parflow are decomposed into subvectors and submatrices.
There is a one-to-one correspondence between the subvectors/submatrices and the
subgrids composing the mesh.
Consequently, they are inherently distributed in parallel via the same rule
\eqref{lexic}.

\subsection{Parallel exchange of information}

In each time step of a simulation, a subset of degrees of freedom (dof) close
to the boundary of a process' subgrid couples to dofs lying on a foreign
process (subgrid) to compute an update to its values.
In \parflow, such a subset is denoted the ``dependent region.''
The dependent region is derived from the stencil, a term which refers to the
computational pattern determined by the numerical ansatz chosen for
discretization.

Given a stencil, \parflow automatically determines the dependent
region. For each subgrid $S$, special routines loop over all
subgrids in the mesh and check which of those are direct
neighbors of $S$ with respect to the processes' partition.
With the data provided by the dependent region,
\parflow is able to determine the source and destination (i.e., sender
and receiver) processes and
the dofs relevant to the MPI messages required to perform vector and matrix
updates.  We will refer to this information as the MPI envelope. \parflow
implements data exchange between two neighboring subgrids via the use of
a ghost layer, that is, an additional strip of artificial dof along the
edges of each subgrid. Hence, data transfered via MPI messages is read
from a subgrid of the sender and written into the ghost layer of the receiver.
The extent of the ghost layer, i.e., the size of the strip in units of dof, is
also defined by the stencil.

The lexicographic ordering principle of the subgrids has a significant impact
on how the information in the MPI envelope is computed. Each
subgrid $S$, as a data structure, stores its triple $(p_x, p_y, p_z)$.
As a consequence, the processes owning the neighbor subgrids to $S$
are located with simple arithmetic. For example, the top neighbor
of $S$ is owned by the process $P_{\mathrm{own}}(p_x,p_y,p_z + 1)$.
The fact that there is exactly one subgrid per process implies that the process
number provides sufficient information to uniquely identify the sender and
receiver of MPI messages.
%
%
%
One downside of this principle is that the number of processes determines
the size of the subgrids.
Essentially, using few processes requires to use a few large subgrids, while
using many processes makes the subgrids fairly small.
Furthermore, imagining to allow for multiple subgrids per process, the
lexicographic ordering will place them in a row along the $x$-axis, leading to
an elongated and thin shape of a process' domain that has a large
surface-to-volume ratio, and thus prompts a larger than optimal message size.

To organize the storage of subgrids, each process holds two arrays of type
subgrid that we will denote by \texttt{allsubs} and \texttt{locsubs},
respectively.
The first one stores pointers to the metadata (such as coordinates in the
process grid, position and resolution) of all subgrids in the grid, and the
second one stores this information only for the subgrids owned by the process
(which is always one in the upstream version).
Hence, the mesh metadata is replicated in every process inside of
\texttt{allsubs}, which leads to a memory usage proportional to $P$ on every
process.
In practice, this disallows runs with more than 32k processes.

\section{Enhancing scalability and speed of \parflow}

%
%


For large scale computations it is imperative that the
mesh storage is strictly distributed.
With the exception of a minimally thin ghost layer on every process' partition
boundary, any data related to the structure of the process-local mesh should be
stored on this process alone.
As pointed out in the previous section, such mesh storage is not implemented in
\parflow's upstream version.
This affects the runtime as well as the total memory usage:
We identified loops proportional to the total number of processes $P$ in the
grid allocation phase and during the determination of the dependent region that
consume roughly 40 minutes on 32k processes (and would need 1h and 20 minutes
on 64k processes, and so forth).
Our proposed solution to enable scalability to $\mathcal O (10^5)$ processes
and more reads as follows.
\begin{enumerate}
	\item Implement a strictly distributed storage of
		\parflow's computational mesh.
	\item Replace loops proportional to the total number of processes
		with constant-size loops.
	\item Allow \parflow to use multiple subgrids per process.
\end{enumerate}
The first two items are essential to enhance the scalability of the code.
We aim to proceed in a minimally invasive way, reusing most of \parflow's
mesh data structures.
This principle may be called \emph{reinterpret instead of rewrite}.
Of course, establishing an optimized non-lexicographic and distributed mesh
layout is a fairly heavy task, which is why we delegate it to a special-purpose
software library described below.
This removes much of the burden from the first item and lets us concentrate on
the second, which requires to audit and modify the code's accesses to mesh data.
The third item serves to decouple the number of subgrids allocated from $P$,
which adds to the flexibility in the setup of simulations.

Let us now describe the tools and algorithmic changes employed to realize these
ideas.

\subsection{The software library \pforest}

Tree based parallel adaptive mesh refinement (AMR) refers to methods in which
the information about the size and position of mesh elements
is maintained within an octree
data structure whose storage is distributed across a parallel computer.
An octree is basically a 1:8 (3D; 1:4 in 2D) tree structure that can
be associated with a recursive refinement scheme where a cube (square)
is subdivided into eight (four) half-size child cubes (squares).
The leaves of the tree either represent the elements of the computational mesh
directly, or can be used to hold other atomic data structures (for example one
subgrid each).
We will refer to the leaves of an octree/quadtree as quadrants.

The canonical domain associated with an octree is a cube (a square in 2D).
When the shape of the domain is more complex, or when it is a rectangle or
brick with an aspect ratio far from unity
(as is the case for most regional subsurface simulations),
it may be advantageous to consider a union of octrees, conveniently
called a forest.

\begin{figure}
	\centering
	\includegraphics[width=0.5\textwidth]{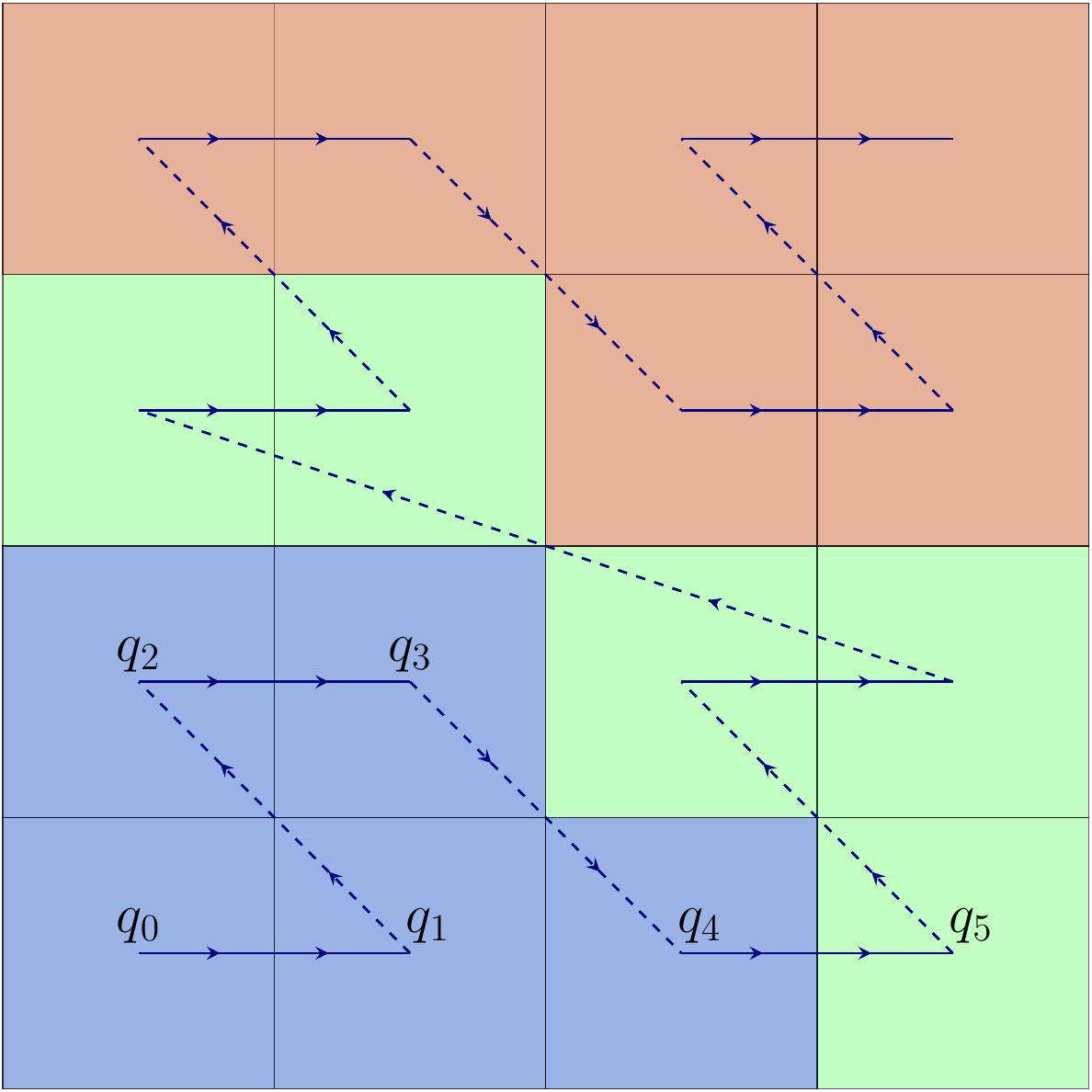}
	\caption{The space filling curve (SFC; zig-zag line) determines an ordering of
	the 16 quadrants $q_i$ obtained by a two-fold subdivision of a square.
        In this example we use the SFC to partition the quadrants between three
        processes (color coded).
        We use dashed lines when two elements that are adjacent in
        the SFC ordering are not direct face neighbors in the domain.
        Since most diagonal lines connect quadrants that are still indirectly
        face-adjacent, the process domains are localized, which makes their
        surface-to-volume ratio less than that of the elongated strip domains
        produced by a lexicographic ordering.
        In fact, it is known that with this SFC each process' subdomain has at
	most two disconnected pieces \cite{BursteddeHolkeIsaac17a}.%
      }
	\label{zcurve}
\end{figure}

The software library
\pforest \cite{BursteddeWilcoxGhattas11,IsaacBursteddeWilcoxEtAl15}
provides efficient algorithms that implement
a self-consistent set of parallel AMR operations.
This library creates and modifies a forest-of-octrees
refinement structure whose storage is distributed using MPI parallelism.
In \pforest a space filling curve (SFC) determines an ordering of the
quadrants that permits fast dynamic re-adaptation and repartitioning;
see Figure \ref{zcurve}.
A \pforest brick structure corresponds to the case in which a forest
consists of multiple tree roots that are arranged to represent a rectangular
Cartesian mesh.

Even when using a uniform refinement and leaving the potential for adaptivi\-ty
unused, as we do in the present work, the
space filling curve paradigm is beneficial since it allows
to drop
the restriction
\eqref{totalprocproduct}:
The total number of processes does not need to match the number of patches used
to split the computational mesh.
Furthermore, using an SFC as opposed to a lexicographic ordering makes each
process' domain more local and approximately sphere-shaped, which
reduces the communication volume on average.

\subsection{Rearranging the mesh layout}

The space filling curve mentioned above provides a suitable
encoding to implement a strictly distributed mesh storage.
Computation of parallel neighborhood relations between processes is part of the
\pforest algorithm bundle and encoded in the \pforest \texttt{ghost} structure.
Mesh generation, distribution and computation of parallel neighborhood
relations are scalable operations in \pforest, in the sense that they
have demonstrated to execute efficiently on parallel machines with
up to 458k processes \cite{IsaacBursteddeWilcoxEtAl15}.

Delegating the mesh management from \parflow to \pforest by identifying each
\pforest quadrant with a \parflow subgrid
allows us to inherit the scalability of \pforest with relatively few
changes to the \parflow code.
In particular, the following features are achievable.
\begin{enumerate}
	\item \parflow's mesh storage can be trimmed down to reference
                only the process-local subgrid(s) and their direct parallel neighbors.
                As a consequence, the memory occupied by mesh storage no longer
                grows with $P$, and loops over subgrids take far less time.
	\item The rule of fixing one subgrid per process can be relaxed.
		This is due to the fact that \pforest has no constraints on
		the number of quadrants assigned to a process.
        \item The computation of \parflow's dependent region is simplified
                by querying neighbor data available through the \pforest
                \texttt{ghost} structure.
\end{enumerate}

The main challenges arising while implementing the identification of a subgrid
with a quadrant are the following.
\begin{enumerate}
	\item Parallel neighborhood relations between processes are
		only known to \pforest. Such information must be correctly passed
		on to the numerical code in \parflow.
	\item The lexicographic ordering of the subgrids will be
		replaced by the ordering established by \pforest via
		the space filling curve.
                This means we must modify the message passing code
                to compute correct neighbor process indices.
	\item We should add support for configurations in which
		a process owns multiple subgrids.
                This will enlarge the range of parallel configurations
                available and enable the option to execute the code on small
                size machines without necessarily reducing the number of
                subgrids employed.
                Additionally, we prepare the code for a subsequent
                implementation of dynamic mesh adaptation, which operates by
                changing the number of subgrids owned by a process at runtime.
\end{enumerate}

The remainder of this section is dedicated to describe how to generate the
\parflow mesh using \pforest.
Essentially, we create a forest of octrees with a specifically computed number
of process-local quadrants and then attach a suitably sized subgrid to each of
them.
Subsequently, we discuss how to obtain information on the parallel mesh layout
from the \pforest \texttt{ghost} interface.

The concept of a fixed number of
processes per coordinate direction is not present in \pforest.
Only the total number of processes is required to compute the process partition
by exploiting the properties of the space filling curve.
Hence, we do not make use of the values of $P_{t}$, $t \in \{x,y,z\}$,
specified by the user.
Instead, we add three variables to the \texttt{tcl} reference script that
provide values for $m_{t}$, the desired numbers of points in a subgrid along
the $t$ coordinate directions.
Then, we rearrange the arithmetic of \eqref{sgdist:a} to compute $P_{t}$ and
$l_{t}$ as derived variables satisfying
\begin{equation}
	\label{newdist:a}
	N_t = m_t \cdot P_t + l_t,
        \quad
	P_t\in\mathbb{N},
        \quad
        l_t\in\{0, \ldots , m_t-1\}
	.
\end{equation}
This construction only interprets $P_t$ as the number of subgrids in the
$t$ direction (while the upstream version of \parflow configures it so).

\begin{figure}
	\centering
	\small
	\includegraphics[width=0.6\textwidth]{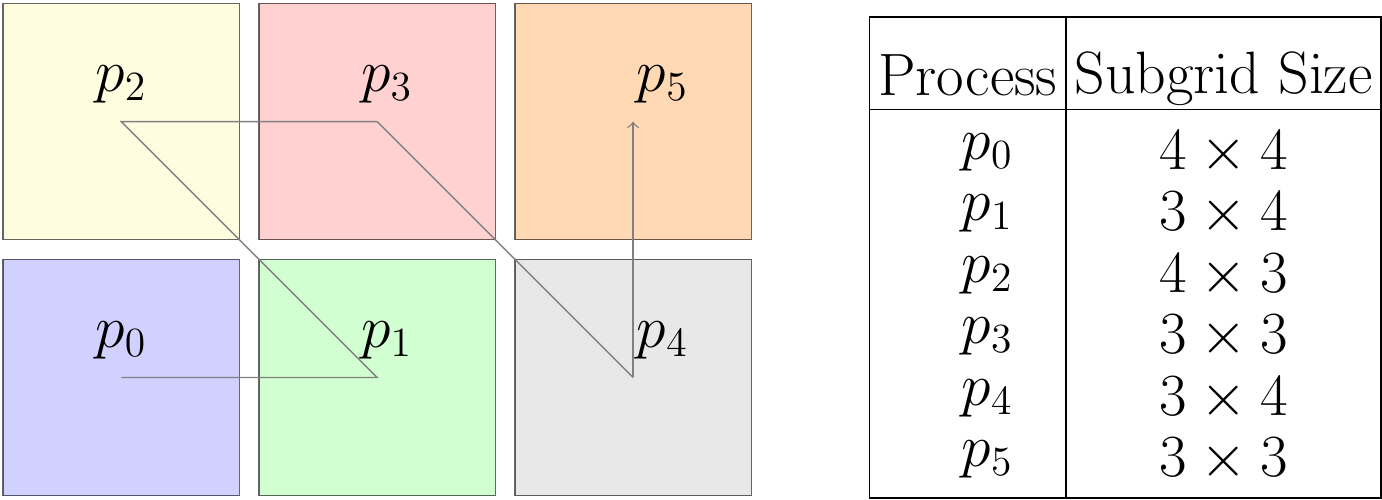}
	\caption{Left, example of a \pforest brick distributed among 6
		processes. Its dimensions are obtained by (\ref{brick:dims})
		after using $N_x=10$, $N_y=7$, $N_z=1$ and $m_{x}=m_{y}=3$, $m_z=1$
		as input values in formula (\ref{newdist:a}).  Right, the size of
		 the subgrids attached to each quadrant according to formula
		 (\ref{sgdist:mt}).
               }
	\label{p4estbrick}
\end{figure}

We must create a \pforest object with $K:=P_{x}\times P_{y}\times P_{z}$
total quadrants.
To do so, we find the smallest box
containing $P_{t}$ quadrants in the $t$ direction and then
refine it accordingly.
Let $k_0$ and $g$ be defined as
\begin{equation}
    \label{brick:k0g}
	k_{0} :=
        \max_{k\in\mathbb{N}} \left\lbrace 2^{k}\mid\gcd(P_{x},P_{y},P_{z})
        \right\rbrace ,
	\qquad
        g:= 2^{k_{0}}
	.
\end{equation}
Thus, $g$ is the biggest power of two dividing the
greatest common divisor of $P_{x}$, $P_{y}$ and $P_{z}$.
Then, a \pforest brick with dimensions
\begin{equation}
	\label{brick:dims}
	P_{x}/g,\quad
	P_{y}/g,\quad
	P_{z}/g
	,
\end{equation}
and refined $k_{0}$ times will have exactly $K$ quadrants.
A brick mesh resulting from applying these rules is shown in
Figure~\ref{p4estbrick}.

\subsection{Attaching subgrids of correct size}

The subgrids must be a partition of the domain in the
sense that their interiors are pairwise disjoint and the union
of all of them cover the grid defined by the user. These conditions
impose restrictions on the choice of the parameters
$N_{t}$ and $m_{t}$.
Specifically, for each $t\in\{x,y,z\}$
we must require
\begin{equation}
\label{condition}
	\sum_{p_t=0}^{P_t - 1} q(p_t) = N_{t},
\end{equation}
in order to satisfy \eqref{newdist:a}.
Recall that $q(p_t)$ is defined in (\ref{sgdist:mt})
as the length in grid points of the $p_t$'th subgrid along direction $t$.
Condition (\ref{condition}) is checked prior the grid allocation
phase, and in case of failure quits the program with a suitable error message
specifying the pair of parameters that violated it.

We inspect the bottom left corner
of each of the quadrants in the \pforest brick, which by construction are only
those that are local to the process, to choose the proper size of the subgrid
that should be attached to it.
In the native \pforest format, each of these corners is encoded by three 32 bit
integers that we scale with the quadrant multiplier $g$ from \eqref{brick:k0g}.
This translates it into integer coordinates that match the enumeration
$p_{t}\in\{0,\ldots P_{t}-1\}$ and are consistent with the rule
(\ref{sgdist:mt}).
Thus, we can use these numbers to determine the position and dimensions of the
\parflow subgrid metadata structure that we allocate and attach to each
quadrant.
%
%

\subsection{Querying the ghost layer}

\begin{figure}
	\centering
	\subfloat[]{
		\includegraphics[width=0.25\textwidth]{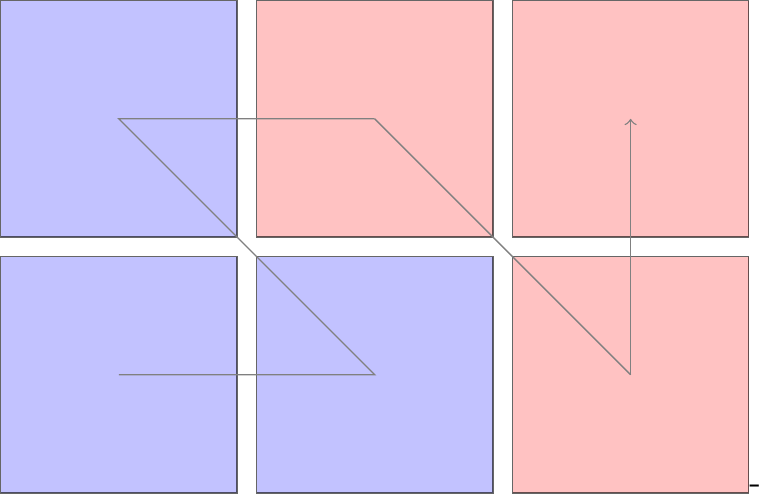}
	}
	\hspace{2ex}
	\subfloat[]{
		\includegraphics[width=0.25\textwidth]{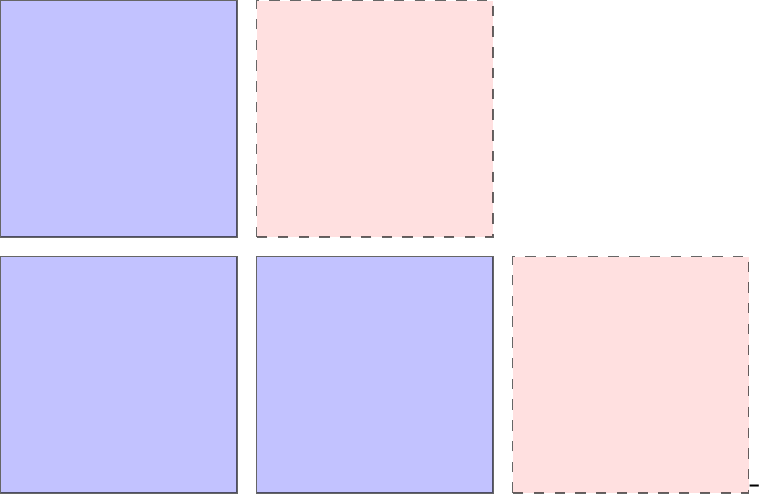}
	}
	\hspace{2ex}
	\subfloat[]{
		\includegraphics[width=0.25\textwidth]{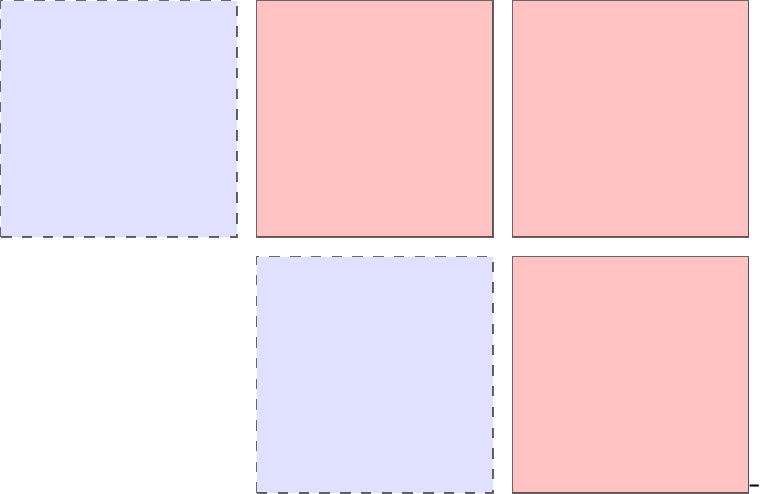}
	}
        \caption{(a) A \pforest brick with six quadrants (identified with
                one subgrid each) distributed among two processes (color
                coded).
                In (b) and (c) we show the view of the brick from
		processes zero and one, respectively.
                The solid boxes represent process-local quadrants,
                for which we allocate mesh metadata and dof storage.
                The dashed boxes represent quadrants in the ghost layer.
                For the ghost quadrants we allocate mesh metadata
                but \emph{no} dof storage.
                Just as the local quadrants, the ghost quadrants are
                traversed in the order of the space filling curve (gray).
              }%
	\label{ghost}%
\end{figure}

We utilize the \texttt{ghost} interface of \pforest to obtain the parallel
neighborhood information required.
Specifically, the ghost data structure provides an array of off-process
quadrants that are direct face neighbors to the local partition; we call these
ghost quadrants (see Figure \ref{ghost}).
We should then populate these quadrants with suitable subgrid metadata that we
use to track their identification on their respective owner processes.
The \pforest \texttt{ghost} object provides the necessary information to do this,
including the lower left corner of each ghost quadrant.
In fact, we can use the enumeration $p_{t}\in\{0,\ldots P_{t}-1\}$
that serves as input to equation (\ref{sgdist:mt}) to compute the
dimensions of both the local and ghost subgrids.
This is most easily done by extending the loop over the \pforest brick
quadrants described above such that it also visits the ghost
quadrants.

While we retain the interpretation of the subgrids array of the upstream
version, storing pointers to the process-local subgrids, we remove almost
all storage in the array \texttt{allsubs}:
The modified version stores pointers to local and ghost subgrids in it, which
are roughly a $1/P$ fraction of the total number of subgrids, and avoids
allocation of even the metadata of all other, locally irrelevant subgrids that
would be of order $P$.
The advantage of this method is that most of the \parflow code does not need to
be changed:
When it loops over the \texttt{allsubs} array, the loops will be radically
shortened, but the relevant logic stays the same.
This is the one most significant change to enable scalability to the full size
of the \juqueen supercomputer (we describe these demonstrations in
Section~\ref{sec:performance} below).

\subsection{Further enhancements}
\label{sec:other_changes}

We have edited \parflow's code for reading configuration files.  If \pforest is
compiled in, we activate the according code at runtime depending on the value
of a new configuration variable.
Hence, even if compiled with \pforest, a user has the flexibility to still use
the upstream version of \parflow on a run-by-run basis.

Additionally, we have written an alternative routine to access and distribute
the information from the user-written configuration file.
As before, the file is read from disk by one process and sent to all other
processes,  We have updated the details of this procedure, since we noticed
that for high process counts (greater equal 65,536) the routine distributed
incorrect data due to an integer overflow, causing the program to crash during
the setup phase.
The modified version delegates this task to the \texttt{MPI\_Bcast} routine,
which works reliably and fast on the usual data size of a few kilobytes.

While running numerical tests, we also encountered some memory issues.
Essentially, memory allocation in \parflow was increasing exponentially with
the number of processes.
With the help of the profiling tool Scalasca \cite{GeimerWolfWylieEtAl10}, we
located the source of the problem in the preconditioner.
Preconditioners in \parflow are managed by the external dependency \hypre
\cite{HypreTeam12}, which is generally known for its scalability.
Since the bug had already been resolved by the \hypre community, an update to
the latest version was sufficient to cure the issue.

\section{Performance evaluation}
\label{sec:performance}

In this section we evaluate the parallel performance of the modified 
version of \parflow.
We follow the concepts of strong and weak scaling studies.
In a strong scaling analysis, a fixed problem is solved on an increasing number
of processing cores and the speedup in runtime is reported.
In a weak scaling study, we increase the problem size and the number of cores
simultaneously such that the work and problem size per core remain the same.
Ideally, the runtime should remain constant for such a study.

Weak and strong scaling studies in this work are assessed on
the massively parallel supercomputer \juqueen.
\juqueen is an
IBM Blue Gene/Q system with 28,672 compute nodes, each with 16 GB of memory
and 16 compute cores, for a total of 458,752 cores.
The machine supports four way simultaneous multi-threading, though we do not
make use of this capability in our studies and always run one process per core.

For each of the experiments, we collect timing information for the
entire simulation.
Additionally, we report timings for
different components of the simulation like the solver setup,
the solver itself and \pforest wrap code executed.
Concerning the wrap code, we refer to additional code written to
execute \pforest interface functions and to retrieve
information from their results and propagate it to \parflow variables.
Particularly important for our purposes are the measurements related to the
solver setup:
The parameters for grid allocation and the management data required for the
parallel exchange of information are computed during this phase.

In the past, parallel scalability of \parflow has been evaluated mainly using
weak scaling studies; see e.g.,
\cite{KolletMaxwell06,KolletMaxwellWoodwardEtAl10,Osei-KuffuorMaxwellWoodward14,GasperGoergenShresthaEtAl14}.
In order to produce unbiased comparisons with these results, which are based on
past upstream versions of \parflow, we keep one subgrid per process in most of
our weak scaling studies, even though we have extended the modified version
to use one or optionally more. We made use of this new feature in
our strong scaling studies; see Figure~\ref{strong_mult}.

\subsection{Weak scaling studies}
\label{sec:weak}

In this section we present results on the weak scalability of the modified
version of \parflow.
We set up a test case in which a global nonlinear problem with integrated
overland flow must be solved. The test case was published previously in
\cite{Maxwell13} and consists of a 3D regular topography problem in which
lateral flow is driven by slopes based on sine and cosine
functions. The problem has a uniform subsurface with space discretization
$\Delta x = \Delta y = 1.0~\mathrm{m}$, $\Delta z = 0.5~\mathrm{m}$.
It is initialized
with a hydrostatic pressure distribution
such that the
top $10.0~\mathrm{m}$ of the aquifer are initially unsa\-turated.
By doubling
the number of grid points $N_x$ and $N_y$, the horizontal extent of the
computational grid is increased by a factor of four per scaling step.
The number of grid points in the vertical direction $N_z$ remains
constant per scaling step. The unit problem has dimensions
$N_x=N_y=50$ and $N_z=40$, meaning that the problem size per process
is fixed to 100,000 grid points. The problem was simulated until
time $t=10.0~\mathrm{s}$ using a uniform time step $\Delta t = 1.0~\mathrm{s}$.

\begin{figure}
	\centering
	\includegraphics[width=0.5\textwidth]{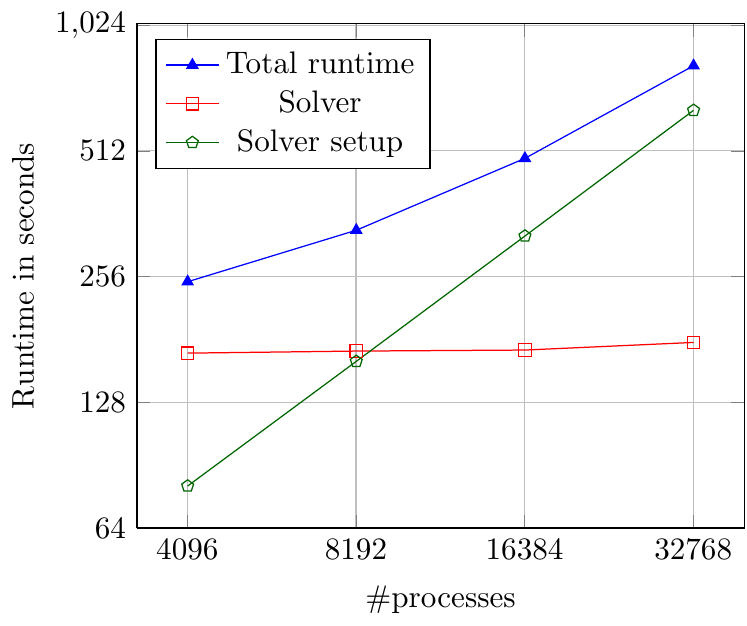}
	\caption{Weak scaling timing results for the upstream
		version of \parflow. 
		The total runtime grows with the number
		processes.
                Breaking it down into solver setup and solver execution shows
                that the former is responsible for the increase of the total
                runtime.
                The solver setup time grows from 81 to 639 seconds, which is
                perfectly proportional to the 8x increase in problem size.}
	\label{weak_no_p4est}
\end{figure}

In order to offer a self-contained comparison with possible improvements
in the \parflow model platform, we conduct the weak scaling study twice, once
with the upstream version of \parflow and then with the modified version with
\pforest enabled.
In the first case, we see that the total runtime grows with the number of
processes.
As can be seen in Figure~\ref{weak_no_p4est}, the solver setup routine is
responsible for this behavior.
Its suboptimal scaling was already reported in
\cite{KolletMaxwellWoodwardEtAl10},
which is in line with us noticing several loops over the all-subgrids array
in this part of the code, which make the runtime effectively proportional to
the total number of processes.

In the experiments we were not able to run the upstream code for 65,536
processes or more, which we attribute to a separate issue in the routine that
reads user input from the configuration reference script, as we detail in
Section~\ref{sec:other_changes}.

\begin{figure}
	\centering
	\includegraphics[width=0.5\textwidth]{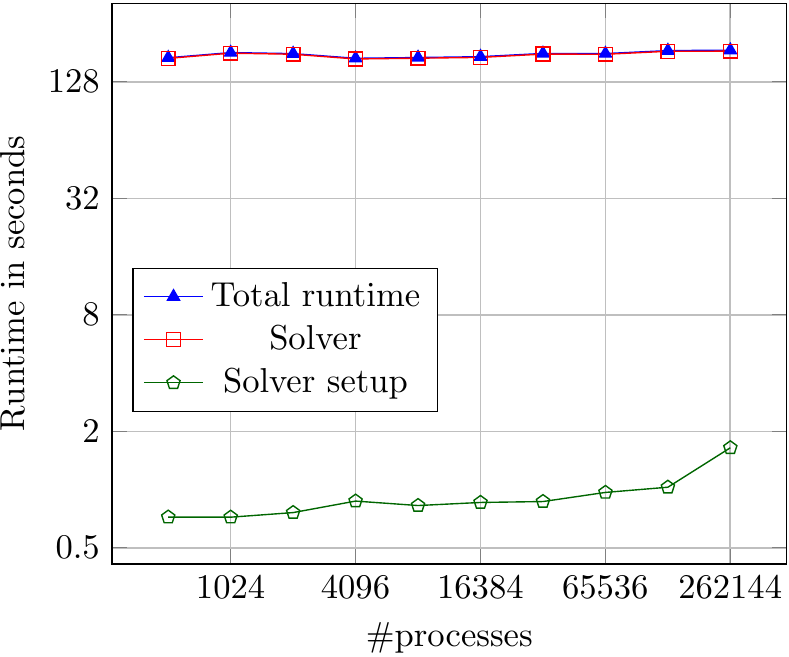}
	\caption{Weak scaling timing results of the modified
		version of \parflow.  The total and solver runtimes
		are nearly identical.
                In comparison to Figure~\ref{weak_no_p4est}, the
		solver setup executes in negligible time,
		ranging between $0.72$ and $1.64$ seconds.}
	\label{weak_p4est}
\end{figure}

\begin{figure}
	\centering
	\subfloat[]{
		\includegraphics[width=0.4\textwidth]{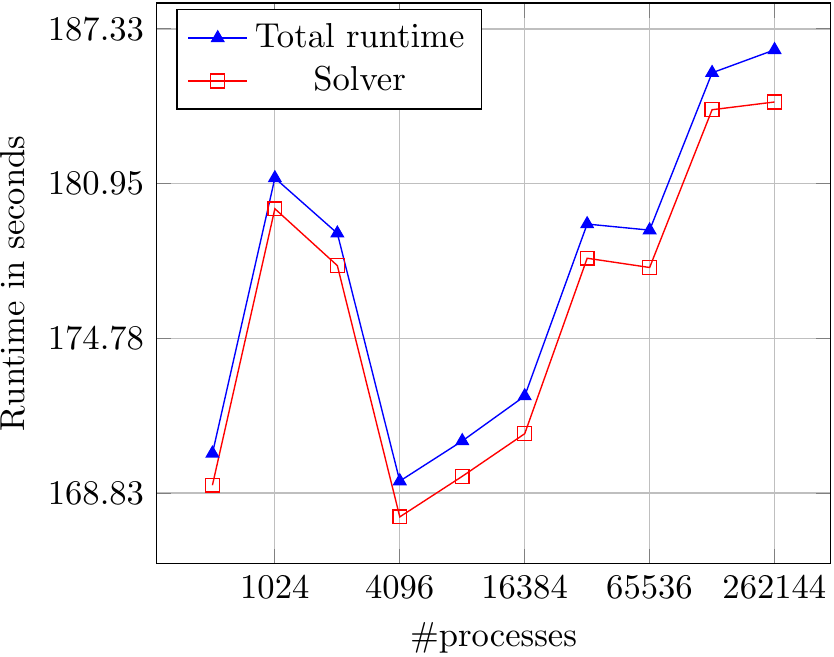}
	}
	\hfill
	\subfloat[]{
		\includegraphics[width=0.4\textwidth]{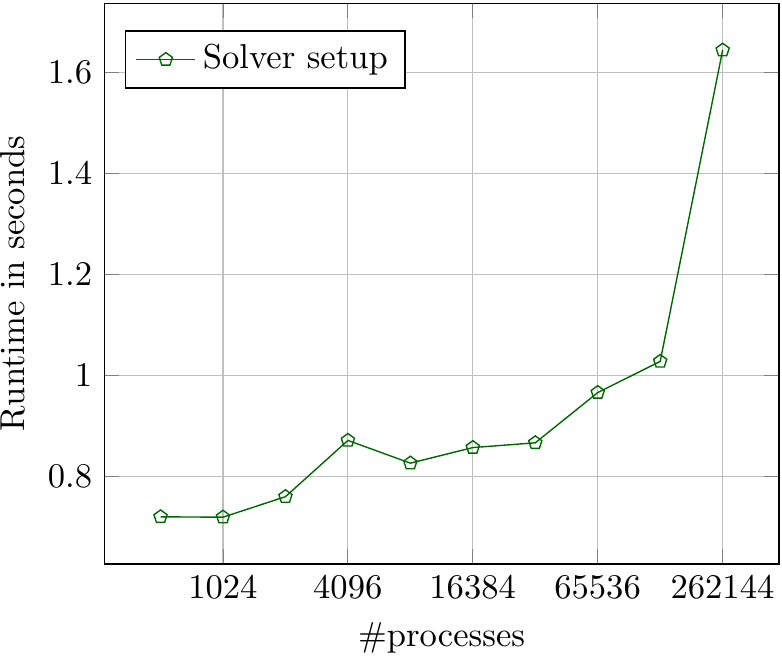}
	}
	\caption{Split of weak scaling timing results from Figure
		\ref{weak_p4est}. In (a) we display the total and solver runtimes
                that vary little in relative terms.
		In (b) we see that the solver setup time
		of the modified version of \parflow stays under two seconds
                wallclock time.}
	\label{weak_p4est_split}
\end{figure}

\begin{figure}
	\centering
	\scriptsize
	\includegraphics[width=0.5\textwidth]{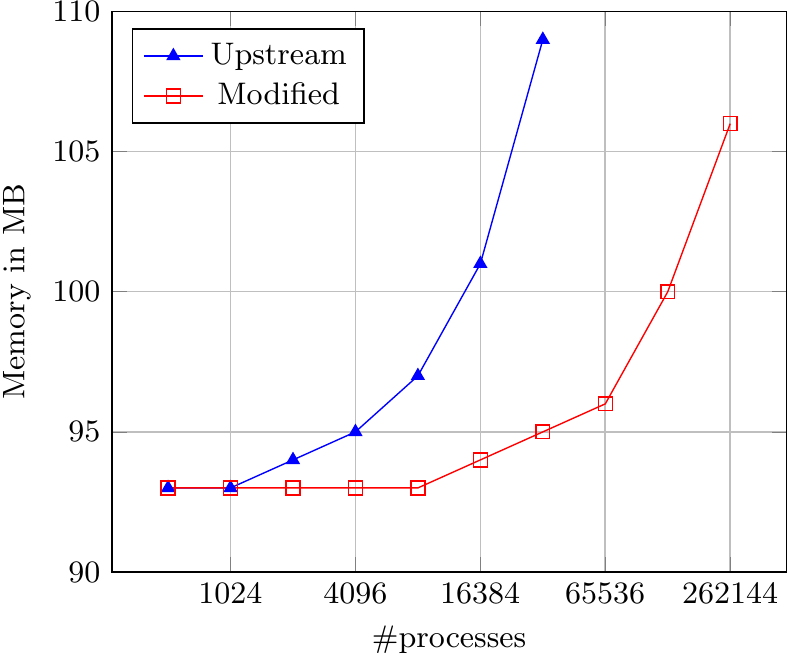}
	\caption{Weak scaling memory usage for the upstream and modified
		versions of \parflow, respectively.
                We record the maximum heap allocation per process
		and plot the maximum of this quantity over all processes.
}
	\label{weak_p4est_mem}
\end{figure}

Repeating the exercise with the modified version of \parflow
dramatically improves the weak scaling behavior of the solver setup;
see Figures~\ref{weak_p4est} and~\ref{weak_p4est_split}.
With \pforest enabled, we replace all loops over the total number of processes
with loops of constant length.
This drops the setup time from over ten minutes at 32k processes to under two
seconds (by a factor of over 300).
Additionally, our patch to the routine that reads the user's configuration
allows us to use as many as 262,144 processes for this study
with nearly optimal, flat weak scaling.
Our implementation of a
strictly distributed storage of \parflow's mesh also leads to a reduction in
memory usage at large scale; see Figure~\ref{weak_p4est_mem}.
This is significant for computers like \juqueen, which may only offer about two
hundred megabytes if the executable is large, especially in combination with
multi-threading.

\begin{figure}
	\centering
	\small
	\subfloat[]{
		\includegraphics[width=0.45\textwidth]{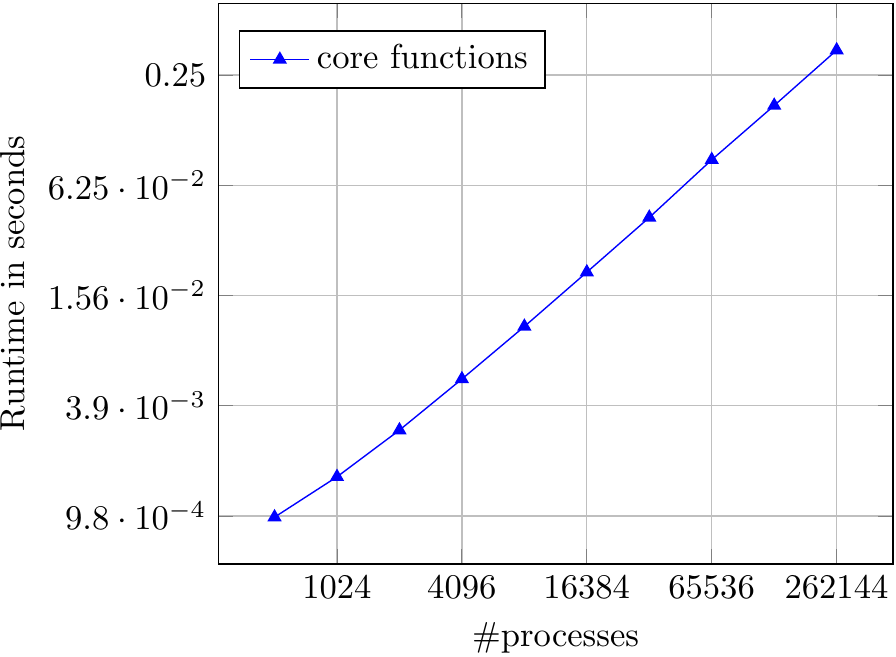}
	}
	\hfill
	\subfloat[]{
		\includegraphics[width=0.45\textwidth]{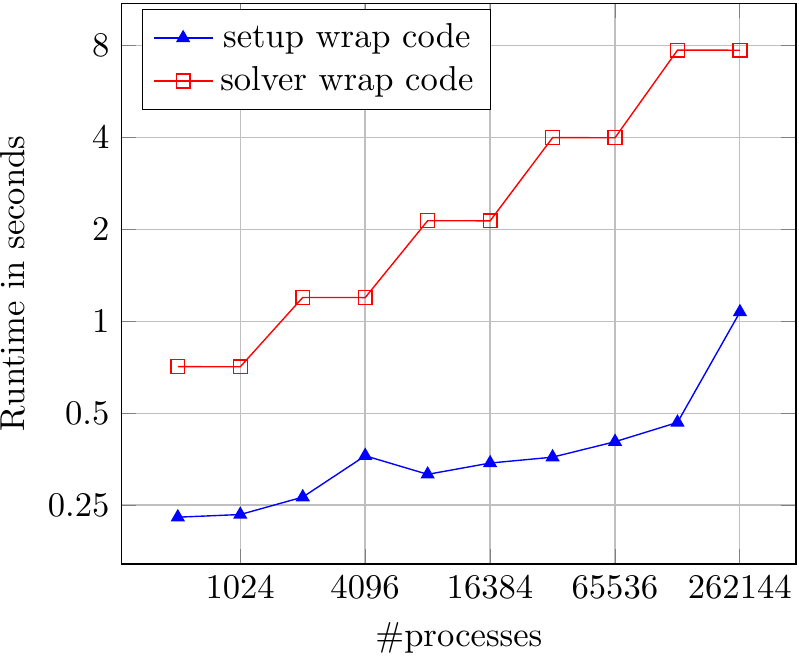}
	}
	\caption{Top: overall timing for \pforest toplevel functions
		used in the code.
                The absolute runtimes are well below one second.
                Bottom: We show timing results for the wrap code executed when
                \pforest is enabled.
                We measure solver and setup related routines independently.
                Compared with the total solver time (see previous Figures), the
                wrap code amounts to a fraction of about 7\% at most.}%
	\label{weak_wrap_code}%
\end{figure}

We also made use of this scaling study
to evaluate the relative cost of using \pforest as new mesh backend by measuring
the overall timing of \pforest related functions introduced in in \parflow.
Our results are displayed in Figure~\ref{weak_wrap_code}.

Additionally, we
employed this test case to estimate the performance of the code in terms of
the floating point operations per second (FLOP/s) and compare them to the
theoretical peak of the Juqueen machine. Measurements were obtained by
instrumenting the code with the Scalasca profiler, which gives access to
the hardware counters from the Performance Application Interface (PAPI)
\cite{PAPI}. In particular, the number of floating point operations from
a whole run have been collected and the FLOP/s estimated as a derived metric
using the CUBE browser \cite{SaviankouKnoblochVisserEtAl15}. We display
our results in Table~\ref{tab:flops}.
\begin{table}
	\renewcommand{\arraystretch}{1.4}
	\centering
	\begin{tabular}{ c|cc }
		 & \multicolumn{2}{c}{FLOP/s} \\
		\# Processes & Upstream & Modified \\ \hline
		$256$ &  $1.223\times10^{8}$ & $1.226\times10^{8}$ \\
		$1024$ & $1.192\times10^{8}$ & $1.194\times10^{8}$ \\
		$4096$ & $1.167\times10^{8}$ & $1.169\times10^{8}$
	\end{tabular}
	\caption{Floating point operations per second (FLOP/s) for the solver
	component of the upstream and modified version of \parflow, respectively.
        The numbers
	are identical up to two significant digits. Both versions of
	\parflow use roughly $3.6\%$ of the theoretical
	peak performance that we expect to get from a single Juqueen process, which
	amounts to $3.2 \times 10^{9}$ FLOP/s.}
	\label{tab:flops}
\end{table}

\subsection{Strong scaling studies}
\label{sec:strong}

In this section, we report our results on the strong scalability of the
modified version of \parflow.
The test case is the same as in the previous section, with the exception of
fixing $N_x$ and $N_y$ while trying a range of process counts.
We divide the results of this study into two categories, depending on whether
we allow for multiple subgrids per process or not.

\begin{figure}
	\centering
	\includegraphics[width=0.5\textwidth]{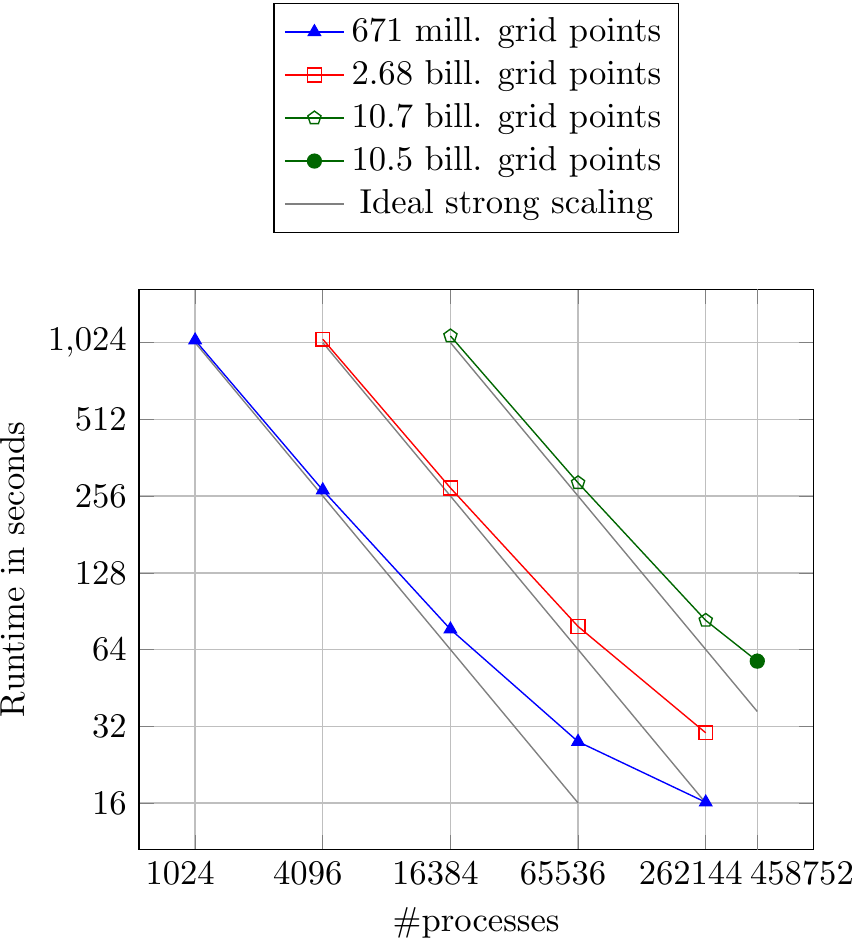}
	\caption{Strong scaling timing results of the modified version of \parflow
	for different problem sizes. We plot the total runtime for
	each case.
        The solid green circle corresponds to the full size of the \juqueen
	system at 458,752 processes.}%
	\label{strong_single}%
\end{figure}

\begin{figure}
	\centering
	\includegraphics[width=0.5\textwidth]{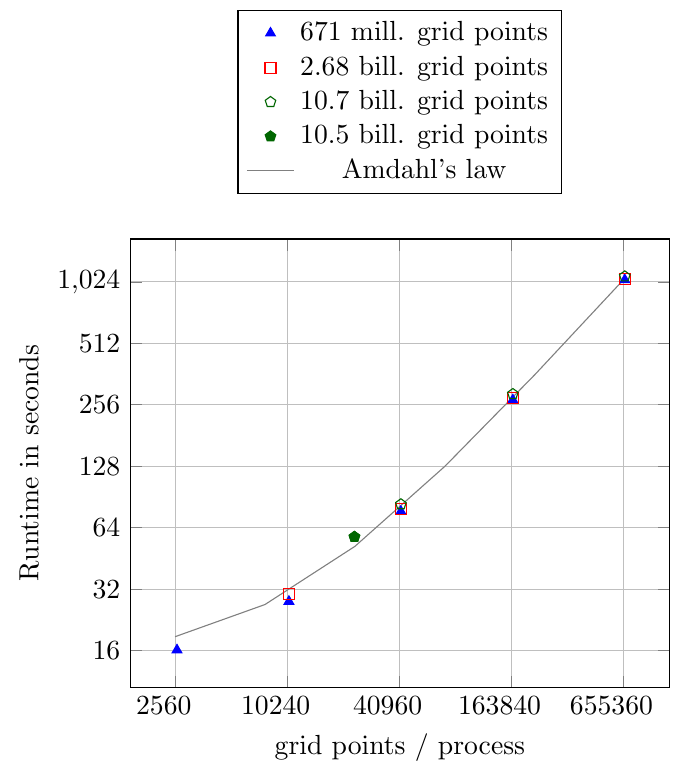}
	\caption{Strong scaling timing results of the modified version of \parflow
		for different problem sizes. We plot the total runtime for
		each case against the number of grid points per process. The solid
		line shows a fit of Amdahl's law $t=c_1 \cdot x + c_2$, where $x$
		denotes the number of grid points per MPI rank. The fitted parameters
		are $c_1 = 0.0016$ and $c_2 = 14.7$.
                This diagram offers another perspective on the optimality of
                weak scaling: We
                observe that measurements for simulations with the same
                number of grid points per process lie on top of each other in
                the vertical.
		}%
	\label{strong_single_rev}%
\end{figure}

We start with one subgrid per process.
In order to keep the problem size fixed
when adding more processes, we adjust the subgrid dimensions properly, i.e., by
decreasing the subgrid sizes in the same proportion as the number of processes
increases.
We run three scaling studies, the smallest of which uses a configuration with
roughly 671~million grid points.
In each subsequent study we increase the problem size by a factor of four.
Hence, the largest problem has around 10.7~billion grid points.
In order to use the full \juqueen system under our self-imposed
restriction of keeping one subgrid per process, while still obtaining runtimes
comparable to the previous scaling studies, we tweak the subgrid dimensions and
number of processes requested in such a way that the resulting problem size
is as close as possible to 10.7~billion.
Table \ref{tab:strong_param} presents a summary of the main parameters defining
these scaling studies;
Figures~\ref{strong_single} and~\ref{strong_single_rev} contain our runtime
results.

\begin{table}
	\centering
	\resizebox{0.8\textwidth}{!}{%
	\begin{tabular}{c|c|c|c}
		& Subgrid size & number of processes & Problem size \\
		& $m_x\times m_y\times m_z$ & $P\times Q\times R$ & $Pm_x\cdot Qm_y\cdot Rm_z$ \\[1ex]\hline
		&&& \\
		& $128\times128\times40$ & $32\times32\times1$ & 671 088 640 \\
		& $64\times64\times40$ & $64\times64\times1$ & 671 088 640 \\
Study 1 & $32\times32\times40$ & $128\times128\times1$ & 671 088 640\\
		& $16\times16\times40$ &  $256\times256\times1$ & 671 088 640 \\
		& $8\times8\times40$   & $512\times512\times1$ & 671 088 640 \\[2ex]\hline
		&&& \\
		& $128\times128\times40$ & $64\times64\times1$ & 2 684 354 560 \\
		& $64\times64\times40$ & $128\times128\times1$ & 2 684 354 560\\
Study 2 & $32\times32\times40$ &  $256\times256\times1$ & 2 684 354 560 \\
		& $16\times16\times40$   & $512\times512\times1$ & 2 684 354 560 \\[2ex]\hline
		&&& \\
		& $128\times128\times40$ & $128\times128\times1$ & 10 737 418 240 \\
Study 3 & $64\times64\times40$ & $256\times256\times1$ & 10 737 418 240\\
		& $32\times32\times40$ &  $512\times512\times1$ & 10 737 418 240 \\[2ex]\hline
		&&& \\
Full system & $18\times32\times40$ &  $896\times512\times1$ & 10 569 646 080\\
    run  &  &   & \\
	\end{tabular}}
	\caption{Relevant parameters for the strong scaling study under the restriction
		of one subgrid per process.  The problem size remains
		constant per scaling study by setting up the subgrid dimensions inversely
		proportional to the number of processes.}
	\label{tab:strong_param}
\end{table}

\begin{figure}
	\centering
	\includegraphics[width=0.5\textwidth]{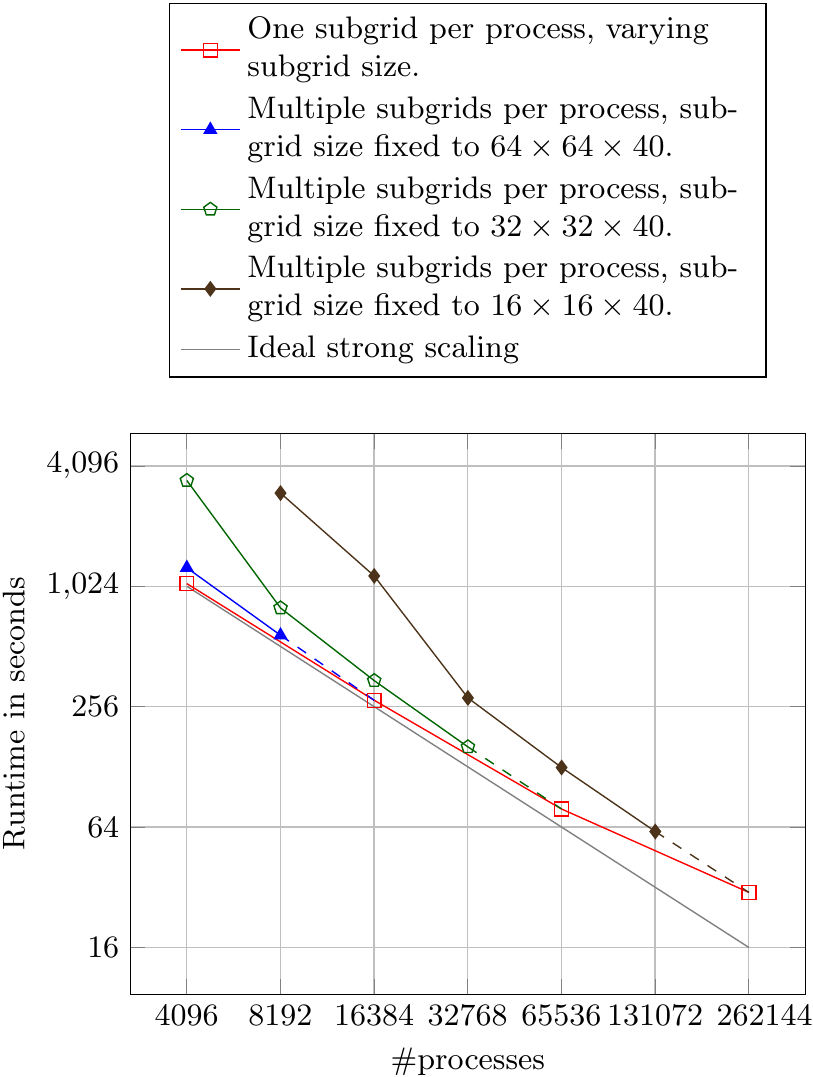}
	\caption{Strong scaling timing results of the modified version of \parflow.
	We compare runs with single and multiple subgrids per process.
        The red line corresponds to the 2.68 billion dof single-subgrid problem
        (also colored red in Figure~\ref{strong_single}).
        Here we choose three fixed subgrid sizes (blue, green, brown).
        Increasing the number of processes eventually leads to a single subgrid
        per process, which is the case already covered in
        Figure~\ref{strong_single}.
        This limit is indicated by the final dotted segment in each graph.
        The largest number of subgrids per process (left end point of each
        graph) is four for the run represented by the blue curve, 16 for the
        green and 32 for the brown.
      }%
	\label{strong_mult}%
\end{figure}
We have designed the modified version of \parflow such that it allows for
configurations in which one process may hold multiple subgrids.
In practice, this option provides additional flexibility when a problem with
a certain size must be run, but the number of processes available depends
on external factors.
Using this feature, we are able to conduct a strong scaling study without
changing the subgrid size with each scaling step.
To illustrate this and additionally to test that the new code supporting such
configurations performs well, we take the medium size problem defined in
Table \ref{tab:strong_param} and execute a classical strong scaling analysis by
changing only the number of processes.
We do this for three different but fixed subgrid sizes.
We present our results in Figure \ref{strong_mult}.
We observe that increasing the number of subgrids per process incurs slightly
higher simulation times but still offers nearly optimal strong scaling
behavior and is thus a viable option compared to the single subgrid
configuration.

\section{Illustrative numerical experiment}
\label{sec:experiment}


In soil hydrology the challenge of scale is ubiquitous.
Heterogeneity
in soil hydraulic properties exists from the sub-centimeter to the kilometer
scale related to, e.g., micro- and macro-porosity and spatially continuous soil
horizons, respectively.
This heterogeneity impacts
the flow and transport
processes in the shallow soil zone and interactions with land surface processes.
Examples are groundwater recharge and leaching of nitrate and pesticides
and the resulting impact on shallow aquifers.
One major structural soil feature is defined by
small scale preferential flow paths, developed from cracking and biota,
that serve as high velocity conduits in the vertical direction.
In large scale simulations, accounting simultaneously for layered soil horizons
and macroporosity at the plot scale on the order to $10^2$ to $10^3$~m has been
essentially impossible,
because of the high spatial resolution required and the enormous size
of the system of equations resulting from the boundary and initial
value problem defined by the 3D Richards equation.

\begin{figure}
	\subfloat[]{
		\includegraphics[width=0.6\textwidth]{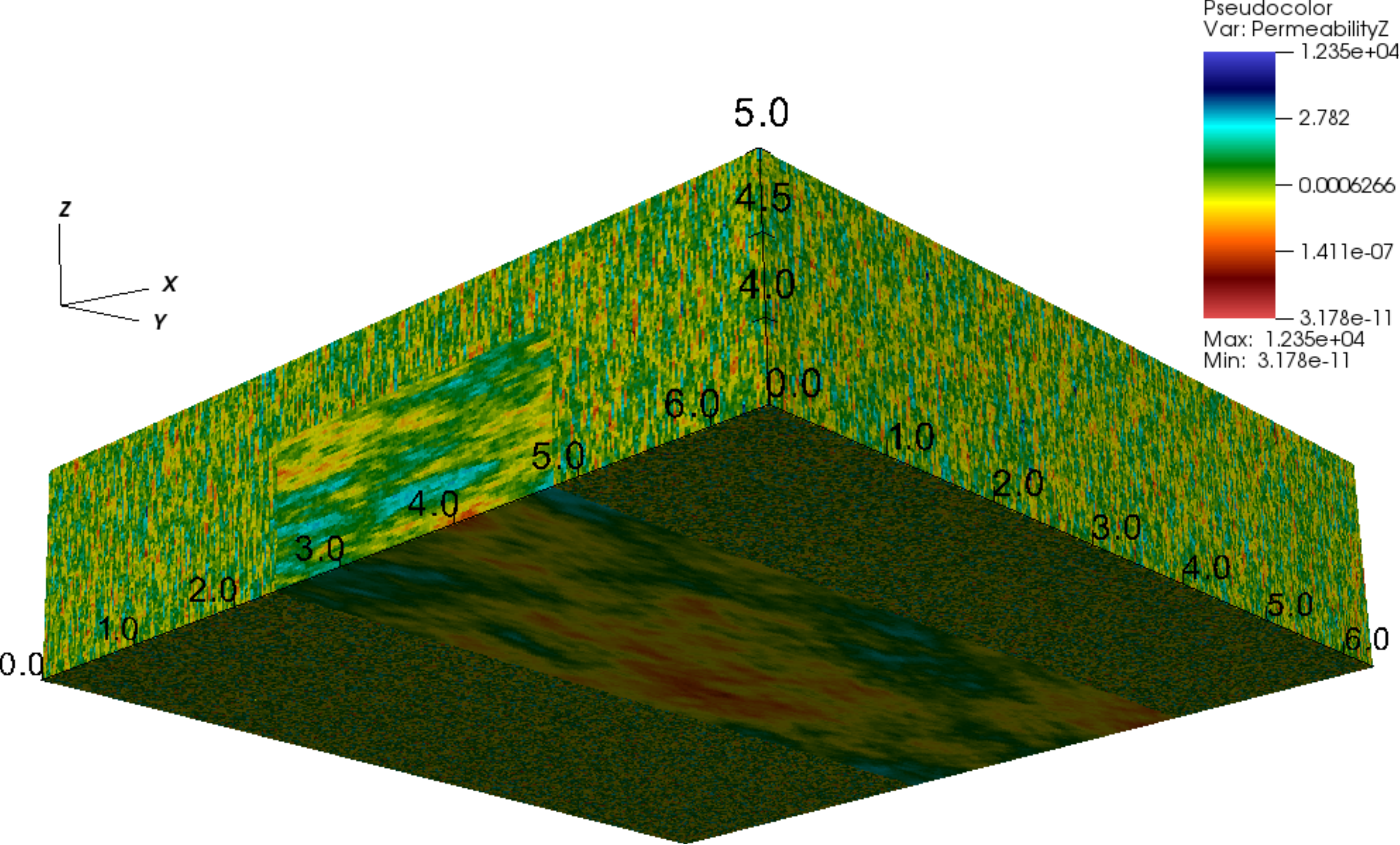}
	}
	\newline
	\subfloat[]{
		\includegraphics[width=0.6\textwidth]{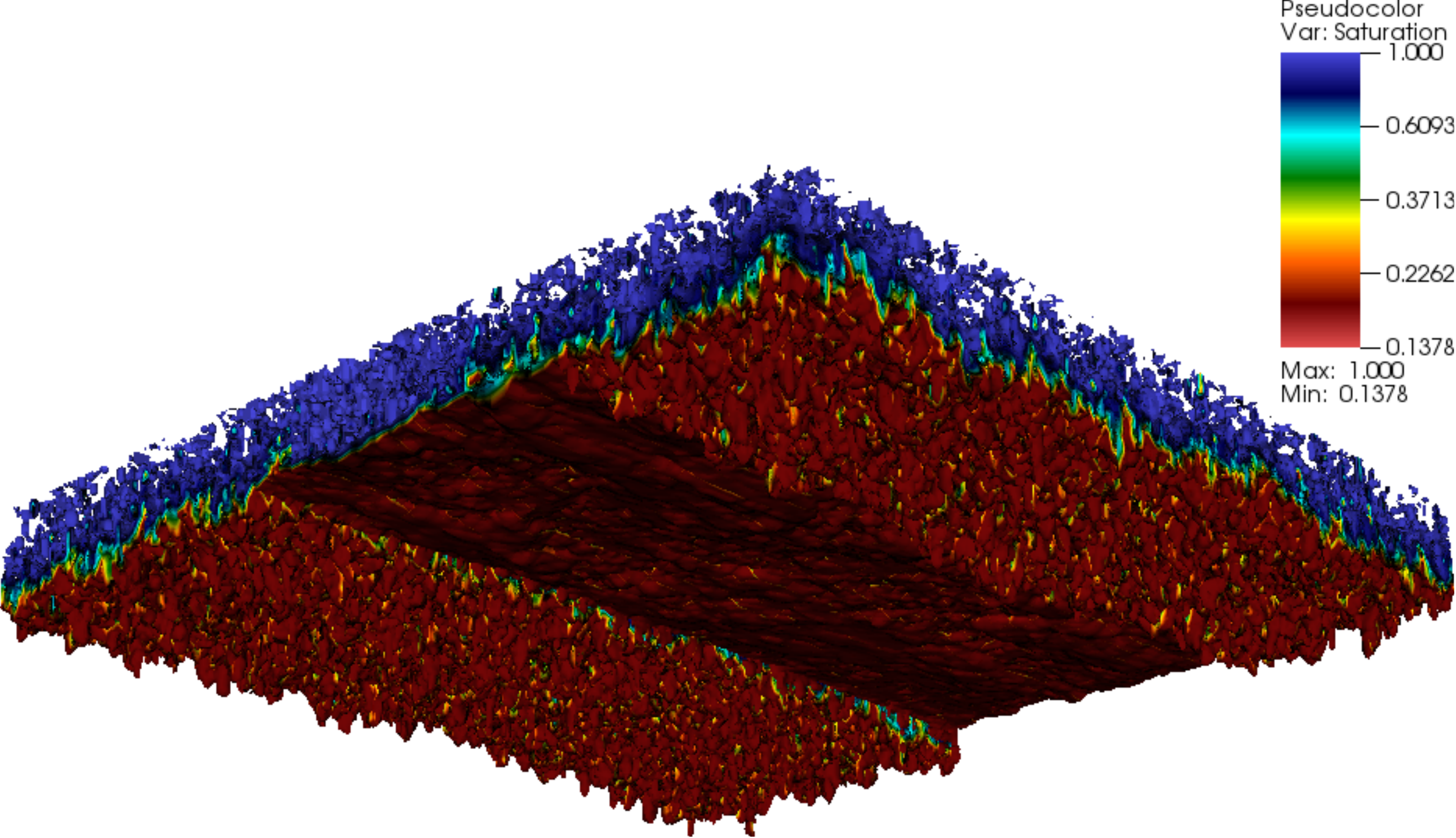}
	}
        \caption{
	Picture a) shows a slice of the permeability field, generated by
	combining two log-transformed Gaussian random fields with
        standard deviations ranging over 3 orders of magnitude.
        In b) we display the saturation field obtained after 32,928 time steps,
        which correspond to 9.83 seconds simulated time.
        The wall clock compute time required for this simulation was around 17
        hours using 16,384 MPI ranks of \juqueen.}
	\label{perm_field}
\end{figure}

The improved parallel performance offered by the modified version of \parflow
motivates us to eventually target such complex simulations.
In order to illustrate this, we simulate a hypothetical example configuration
focused on the presence of macroporosity and layered soil horizons.
The numerical experiment chosen solves an infiltration problem on
a Cartesian domain. The initial water table is implemented as a constant head
boundary at the bottom of the domain with a five meter unsaturated zone on top
of it. The heterogeneous permeability parameter is simulated with a spatially
correlated log-transformed Gaussian random field. We employ two realizations
of such a field to model vertical and lateral preferential flow paths,
respectively. Both random field realizations are
obtained with a parallel random field generator implemented in \parflow (the
turning-bands algorithm \cite{ThomsonAbabouGelhar89}).
We display an example of the outcome of such realizations and the saturation
field obtained from our experiment in Figure~\ref{perm_field}.

The total compute time required was roughly 280,000 core hours.
While the modified version of \parflow can be scaled easily to use large
multiples of this number, using such amounts of time must be carefully
justified.
At this point, it makes sense to reserve extended scientific studies for a
later publication that will focus exclusively on the design and usefulness of
such simulations.

\section{Conclusions}






 




The purpose of this work is to improve the parallel performance of
the subsurface simulator \parflow such it can take full advantage
of the computational resources offered by the latest HPC systems.
Our approach is to couple
the \parflow and \pforest libraries such that the latter acts as the
parallel mesh manager of the former.
We achieve this with relatively small and local changes to \parflow that
constitute a reinterpretation rather than a redesign.
This modified version of \parflow offers a wider range of runnable
configurations and improved performace.
We report good weak and strong scaling up to 458,752 MPI ranks on the \juqueen
supercomputer.

The improved performance of the modified version of \parflow opens
the possibility of bigger and more realistic simulations.
For example, the code can be used to perform virtual soil column experiments to
upscale hydraulic parameters related to hydraulic conductivity and the soil
water retention curve.
One could envision a hierarchy of experiments covering heterogeneities starting
from the laboratory scale up to some 100~m.
The upscaled parameters may then be used in coarser resolution models.

Considering our ehancenments, most parallel bottlenecks are gone and
implementation scalability is in principle unlimited.
Users should be aware that with the modified version of \parflow,
one can easily spend millions of compute hours, which demands
care and sensibility in choosing the experimental setup.
Nevertheless, a certain ratio of losses will be unavoidable when designing
simulations at highly resolved scales due to the process of (informed) trial
and error.

We also note that we did not address the algorithmic efficiency of the time
stepper or the preconditioner, since the mathematics of the solver remain
unchanged.
Future developments in this regard will automatically inherit and benefit from
the improvements in scalability reported here.

We are providing all code changes to the community, hoping for the best
possible use and feedback, and will consider extending the current capabilities
further when needed.

\section*{Acknowledgments}

The development of this work is made possible via the financial support by the
collaborative research initiative SFB/TR32 ``Patterns in
Soil-Vegetation-Atmosphere Systems: Monitoring, Modeling, and Data
Assimilation,'' project D8,
funded by the Deutsche Forschungsgemeinschaft (DFG).  Authors
B.\ and F.\ gratefully acknowledge additional travel support by the Bonn
Hausdorff Centre for Mathematics (HCM) also funded by the DFG.

We also would like to thank the Gauss Centre for Supercomputing (GCS) for
providing computing time through the John Von Neumann  Institute for Computing
(NIC) on the GCS share of the supercomputer \juqueen at the J\"ulich
Supercomputing Centre (JSC).  GCS is the alliance of the three national
supercomputing centers HLRS (Universit\"at Stuttgart),
JSC (For\-schungs\-zentrum J\"ulich), and LRZ (Bayerische Akademie der Wissenschaften), funded by the
German Federal Ministry of Education and Research (BMBF) and the German State
Ministries for Research of Baden-W\"urttemberg (MWK), Bayern (StMWFK), and
Nordrhein-Westfalen (MIWF).

Our contributions to the \parflow code and the scripts defining the
test configurations for the numerical experiments exposed in this work are
available as open source at \url{https://github.com/parflow}.

\bibliographystyle{siam}
\bibliography{./group,./ccgo_new}

\end{document}